\documentstyle[12pt]{article}

\def\journal{\topmargin .3in	\oddsidemargin .5in
	\headheight 0pt	\headsep 0pt
	\textwidth 5.625in % 1.2 preprint size  %6.5in
	\textheight 8.25in % 1.2 preprint size 9in
	\marginparwidth 1.5in
	\parindent 2em
	\parskip .5ex plus .1ex		\jot = 1.5ex}
\journal

\catcode`\@=11
\def\marginnote#1{}
\newcount\hour
\newcount\minute
\newtoks\amorpm
\hour=\time\divide\hour by60
\minute=\time{\multiply\hour by60 \global\advance\minute by-\hour}
\edef\standardtime{{\ifnum\hour<12 \global\amorpm={am}%
	\else\global\amorpm={pm}\advance\hour by-12 \fi
	\ifnum\hour=0 \hour=12 \fi
	\number\hour:\ifnum\minute<10 0\fi\number\minute\the\amorpm}}
\edef\militarytime{\number\hour:\ifnum\minute<10 0\fi\number\minute}
\def\draftlabel#1{{\@bsphack\if@filesw {\let\thepage\relax
   \xdef\@gtempa{\write\@auxout{\string
      \newlabel{#1}{{\@currentlabel}{\thepage}}}}}\@gtempa
   \if@nobreak \ifvmode\nobreak\fi\fi\fi\@esphack}
	\gdef\@eqnlabel{#1}}
\def\@eqnlabel{}
\def\@vacuum{}
\def\draftmarginnote#1{\marginpar{\raggedright\scriptsize\tt#1}}
\def\draft{\oddsidemargin -.5truein
	\def\@oddfoot{\sl preliminary draft \hfil
	\rm\thepage\hfil\sl\today\quad\militarytime}
	\let\@evenfoot\@oddfoot	\overfullrule 3pt
	\let\label=\draftlabel
	\let\marginnote=\draftmarginnote
   \def\@eqnnum{(\theequation)\rlap{\kern\marginparsep\tt\@eqnlabel}%
\global\let\@eqnlabel\@vacuum}  }
\def\preprint{\twocolumn\sloppy\flushbottom\parindent 2em
	\leftmargini 2em\leftmarginv .5em\leftmarginvi .5em
	\oddsidemargin -.5in	\evensidemargin -.5in
	\columnsep .4in	\footheight 0pt
	\textwidth 10in	\topmargin  -.4in
	\headheight 12pt \topskip .4in
	\textheight 7.1in \footskip 0pt
	\def\@oddhead{\thepage\hfil\addtocounter{page}{1}\thepage}
	\let\@evenhead\@oddhead	\def\@oddfoot{}	\def\@evenfoot{} }
\def\numberbysection{\@addtoreset{equation}{section}
	\def\theequation{\thesection.\arabic{equation}}}
\def\underline#1{\relax\ifmmode\@@underline#1\else
	$\@@underline{\hbox{#1}}$\relax\fi}
\def\titlepage{\@restonecolfalse\if@twocolumn\@restonecoltrue\onecolumn
     \else \newpage \fi \thispagestyle{empty}\c@page\z@
	\def\thefootnote{\fnsymbol{footnote}} }
\def\endtitlepage{\if@restonecol\twocolumn \else \newpage \fi
	\def\thefootnote{\arabic{footnote}}
	\setcounter{footnote}{0}}  %\c@footnote\z@ }
\catcode`@=12
\relax
\def\figcap{\section*{Figure Captions\markboth
	{FIGURECAPTIONS}{FIGURECAPTIONS}}\list
	{Figure \arabic{enumi}:\hfill}{\settowidth\labelwidth{Figure 999:}
	\leftmargin\labelwidth
	\advance\leftmargin\labelsep\usecounter{enumi}}}
 \relax
\def\tablecap{\section*{Table Captions\markboth
	{TABLECAPTIONS}{TABLECAPTIONS}}\list
	{Table \arabic{enumi}:\hfill}{\settowidth\labelwidth{Table 999:}
	\leftmargin\labelwidth
	\advance\leftmargin\labelsep\usecounter{enumi}}}
 \relax
\def\reflist{\section*{References\markboth
	{REFLIST}{REFLIST}}\list
	{[\arabic{enumi}]\hfill}{\settowidth\labelwidth{[999]}
	\leftmargin\labelwidth
	\advance\leftmargin\labelsep\usecounter{enumi}}}
 \relax
\makeatletter
\newcounter{pubctr}
\def\publist{\@ifnextchar[{\@publist}{\@@publist}}
\def\@publist[#1]{\list
	{[\arabic{pubctr}]\hfill}{\settowidth\labelwidth{[999]}
	\leftmargin\labelwidth
	\advance\leftmargin\labelsep
	\@nmbrlisttrue\def\@listctr{pubctr}
	\setcounter{pubctr}{#1}\addtocounter{pubctr}{-1}}}
\def\@@publist{\list
	{[\arabic{pubctr}]\hfill}{\settowidth\labelwidth{[999]}
	\leftmargin\labelwidth
	\advance\leftmargin\labelsep
	\@nmbrlisttrue\def\@listctr{pubctr}}}
 \relax
\makeatother
\catcode`\@=11
\def\section{\@startsection {section}{1}{0pt}{-3.5ex plus -1ex minus
 -.2ex}{2.3ex plus .2ex}{\raggedright\large\bf}}
\catcode`\@=12
\newskip\humongous \humongous=0pt plus 1000pt minus 1000pt
\def\caja{\mathsurround=0pt}

\newif\ifdtup
\def\panorama{\global\dtuptrue \openup1\jot \caja
	\everycr{\noalign{\ifdtup \global\dtupfalse
	\vskip-\lineskiplimit \vskip\normallineskiplimit
	\else \penalty\interdisplaylinepenalty \fi}}}
\def\eqalignno#1{\panorama \tabskip=\humongous
	\halign to\displaywidth{\hfil$\displaystyle{##}$
	\tabskip=0pt&$\displaystyle{{}##}$\hfil
	\tabskip=\humongous&\llap{$##$}\tabskip=0pt
	\crcr#1\crcr}}
\def\oldreffmt#1{\rlap{[#1]} \hbox to 2\parindent{}}

\def\figfmt#1{\rlap{Figure {#1}} \hbox to 1in{}}

\def\beq{\begin{equation}}
\def\eeq{\end{equation}}

\def\bea{\begin{eqnarray}}
\def\eea{\end{eqnarray}}
\catcode`\@=11
\def\eqnarray{\stepcounter{equation}\let\@currentlabel=\theequation
\global\@eqnswtrue
\global\@eqcnt\z@\tabskip\@centering\let\\=\@eqncr
\gdef\@@fix{}\def\eqno##1{\gdef\@@fix{##1}}%
$$\halign to \displaywidth\bgroup\@eqnsel\hskip\@centering
  $\displaystyle\tabskip\z@{##}$&\global\@eqcnt\@ne
  \hskip 2\arraycolsep \hfil${##}$\hfil
  &\global\@eqcnt\tw@ \hskip 2\arraycolsep $\displaystyle\tabskip\z@{##}$\hfil
   \tabskip\@centering&\llap{##}\tabskip\z@\cr}

\def\@@eqncr{\let\@tempa\relax
    \ifcase\@eqcnt \def\@tempa{& & &}\or \def\@tempa{& &}
      \else \def\@tempa{&}\fi
     \@tempa \if@eqnsw\@eqnnum\stepcounter{equation}\else\@@fix\gdef\@@fix{}\fi
     \global\@eqnswtrue\global\@eqcnt\z@\cr}

\catcode`\@=12
\font\tenbifull=cmmib10 % bold math italic
\font\tenbimed=cmmib10 scaled 800
\font\tenbismall=cmmib10 scaled 666
\def\boldzeta{\fam=9{\mathchar"7110 } }

\def\boldlambda{\fam=9{\mathchar"7115 } }

\relax

\def\thefootnote{\fnsymbol{footnote}}
\begin{document}
\begin{titlepage}
\begin{center}
\today     \hfill    LBL-36269 \\
          \hfill    UCB-PTH-94/25 \\

\vskip .25in

{\large \bf Electric Dipole Moments as A Test of Supersymmetric Unification}
%\footnote{This work was supported by the Director, Office of Energy
%Research, Office of High Energy and Nuclear Physics, Division of High
%Energy Physics of the U.S. Department of Energy under Contract
%DE-AC03-76SF00098.}
%\vskip .50in
%alternate footnote for faculty:
\footnote{This work was supported in part by the Director, Office of
Energy Research, Office of High Energy and Nuclear Physics, Division of
High Energy Physics of the U.S. Department of Energy under Contract
DE-AC03-76SF00098 and in part by the National Science Foundation under
grants PHY-90-21139 and PHY-92-19345.}

\vskip .25in
S. Dimopoulos \footnote{On leave of absence from Stanford University, Stanford,
CA 94305, USA.}

{\em Theory Division, CERN\\
CH-1211,
Geneva 23, Switzerland.}
\vskip .25in
L.J. Hall

{\em Theoretical Physics Group\\
    Lawrence Berkeley Laboratory\\
and\\
Department of Physics, University of California\\
    Berkeley, California 94720}
\end{center}

\vskip .4in

\begin{abstract}

In a class of supersymmetric grand unified theories,
including those based on the gauge group $SO(10)$,
there are new contributions to the electric dipole moments of the neutron and
electron,
which arise as a heavy top quark effect.
These contributions arise from CKM-like phases, not from phases of the
supersymmetry breaking operators,
and can be reliably computed in
terms of the parameters of the weak scale supersymmetric theory.
For the expected ranges of these parameters,
the electric dipole moments of the neutron and the electron are predicted to be
close to present experimental limits.

\end{abstract}
\end{titlepage}
%THIS PAGE (PAGE ii) CONTAINS THE LBL DISCLAIMER
%TEXT SHOULD BEGIN ON NEXT PAGE (PAGE 1)
\renewcommand{\thepage}{\roman{page}}
\setcounter{page}{2}
\mbox{ }

\vskip 1in

\begin{center}
{\bf Disclaimer}
\end{center}

\vskip .2in

\begin{scriptsize}
\begin{quotation}
This document was prepared as an account of work sponsored by the United
States Government. While this document is believed to contain correct
 information, neither the United States Government nor any agency
thereof, nor The Regents of the University of California, nor any of their
employees, makes any warranty, express or implied, or assumes any legal
liability or responsibility for the accuracy, completeness, or usefulness
of any information, apparatus, product, or process disclosed, or represents
that its use would not infringe privately owned rights.  Reference herein
to any specific commercial products process, or service by its trade name,
trademark, manufacturer, or otherwise, does not necessarily constitute or
imply its endorsement, recommendation, or favoring by the United States
Government or any agency thereof, or The Regents of the University of
California.  The views and opinions of authors expressed herein do not
necessarily state or reflect those of the United States Government or any
agency thereof or The Regents of the University of California and shall
not be used for advertising or product endorsement purposes.
\end{quotation}
\end{scriptsize}

\vskip 2in

\begin{center}
\begin{small}
{\it Lawrence Berkeley Laboratory is an equal opportunity employer.}
\end{small}
\end{center}

\newpage
\renewcommand{\thepage}{\arabic{page}}
\setcounter{page}{1}
%THIS IS PAGE 1 (INSERT TEXT OF REPORT HERE)

{\bf 1.}

Grand unification is based on the idea that, at a fundamental level,
quarks and leptons are identical and cannot be distinguished [1].
This simple idea of a symmetry which relates quarks to leptons has
several attractive features.
For example, in $SO(10)$ theories [2], the quarks and leptons of a single
generation are the components of a single spinor representation.
The particle physics version of the periodic table, the $SU(3) \times SU(2)
\times U(1)$ gauge quantum numbers of a generation, are immediately understood
in terms of the group properties of this spinor representation.
An indication that grand unification may occur at very high energies is
provided
by the unification of the gauge coupling constants [3], which gives agreement
with the experimental value of the weak mixing angle at the present level of
accuracy, providing supersymmetry is present at the weak scale [4].
Low energy supersymmetry is itself an aid in attacking the hierarchy problem,
and can lead to a dynamical breaking of $SU(2)\times U(1)$ [5], thus linking
the
$Z$ boson mass to the scale of supersymmetry breaking.

While this picture of supersymmetric grand unification appears plausible, the
mass scale of the new unified interactions, $M_G$, is enormous: $M_G \approx
10^{16}$ GeV.
How can such a theory be tested?

If the weak scale theory is the standard model, then effects from the grand
unified theory can appear in only two ways.
Once the particle content of the standard model is fixed, the renormalizable
interactions of the standard model are the most general which are consistent
with the $SU(3) \times SU(2) \times U(1)$ gauge symmetry.
Hence the grand unified theory can either provide relationships between the
free
couplings of the standard model, or it can induce additional non-renormalizable
interactions.
There are no other possibilities.
The weak mixing angle prediction provides a well known example of the former
effect.
The prediction for $m_b/m_\tau$ [6] provides a second example, in fact up to
seven predictions are possible in the flavor sector of SO(10) theories [7].
However, the weak mixing angle prediction is unique in its conceptual
simplicity, it requires only that the unified group contains $SU(5)$, whereas
the flavor predictions require further assumptions.

Since the non-renormalizable interactions give effects suppressed by powers of
$1/M_G$, the signals must be ones which are absent or highly suppressed in the
standard model.
There are only two such signals: those which violate baryon
numbers ($B$), such as proton decay, and those which violate lepton member
($L$),
such as neutrino masses.

In supersymmetric grand unified theories it is not easy to reliably calculate
the $B$ and $L$ violating effects from the non-renormalizable interactions.
The simplest estimates give neutrino masses too small to observe in terrestrial
experiments, and a proton decay rate which is already excluded.
More refined calculations can lead to interesting rates, but are subject
to other difficulties.
For example, the rates depend on superheavy particle masses and interactions
which are frequently unknown, and are certainly model dependent.
Furthermore, there is no compelling minimal supersymmetric unified model.

However, in supersymmetric theories there is a third way in which the high
energy interactions can be manifest at low energies [8].
As well as relating renormalizable parameters of the low energy supersymmetric
theory and inducing non-renormalizable
operators, the unified interactions can change the form of the soft
supersymmetry breaking interactions.
This window to the high energy domain is only open if the original form of the
soft
supersymmetry breaking interactions is not the most general allowed by the low
energy symmetries, and if the supersymmetry breaking effects are present at or
above the scale $M_G$.
We now discuss these two points.

Although supersymmetry can break in a great many ways, for our proposes the
only
crucial question is whether the superpartners first feel the effects of
supersymmetry breaking above or below the scale $M_G$.
For example, if supersymmetry is broken dynamically by some new force at scale
 $\Lambda \ll M_G$, and if this breaking is communicated to the superpartners
by
 gauge interactions at this scale, then the grand unified interactions can only
 modify the form of the supersymmetry breaking by effects which are suppressed
 by powers of $\Lambda/M_G$.
The third window to the unified interactions is closed.
On the other hand, if the soft supersymmetry breaking interactions are present
at the scale $M_G$, they can be directly modified by the
unified interactions and the window is open.
An example of the latter case is when the supersymmetry breaking occurs at the
intermediate           scale, and is communicated to the superpartners by
supergravitational interactions [9].
In this case the soft operators for the superpartners are present up to the
Planck scale, the situation assumed in this letter.

What form do the soft operators have at the Planck scale?
If the form is the most general allowed by
the gauge symmetry, then flavor changing processes [10]
and the neutron electric dipole moment [11] force the scale of supersymmetry
breaking to be unnaturally far above the weak scale.
Hence in this letter we follow the common practice of assuming a boundary
condition for the soft operators which does not distinguish generations and
which conserves CP.
The generation and CP dependence of the soft operators at low energies then
reflects the interactions of the
grand unified theory.\footnote{ It may be objected that if we see such
flavor changing and CP violating effects we cannot be sure that they originate
from the grand unified interactions rather than from small effects at the
Planck scale in the boundary conditions.
While this is clearly true, the crucial difference is that we know how to
reliably compute the grand unified effects, whereas it is not known how to
compute small generation dependent or CP violating terms at the Planck scale.}

The crucial imprint which the high energy interactions leave on the soft
operators [8] can be seen from the following simple example.
Consider an $SO(10)$ unified model in which the 16-plet of the ith generation,
$16_i$, has a coupling $\lambda_i 16_i AB$ where $A$ and $B$ are any fields,
subject only to the condition that at least one of them is superheavy with a
mass $M_G$.
This interactions provides a 1-loop radiative correction to the soft
mass-squared matrix for the superpartners in the $16_i$
$$
{\triangle m^2_{ij}\over m^2_0} \approx {\lambda_i\lambda^*_j\over \pi^2} \ln
{M_P\over M_G}\eqno(1)
$$
where $m_0$ is the common scalar mass at the Planck mass $M_P$.
The choice of normalization, $1/\pi^2$, is justified by the precise equations
which follow.
Thus we see that $\lambda_i = 1$ leads to $\approx 50\%$ corrections in the
eigenvalues of the soft masses.
Furthermore, $\lambda_i \neq \lambda_j$ leads to flavor and CP violation.
Thus, not only do the effects of the interactions of superheavy particles not
decouple with powers of $1/M_G$, but they lead to enormous effects at the weak
scale if the coupling constants have strength unity.
As a further advantage, there is only a logarithmic sensitivity to the mass of
the superheavy particle, which will not be precisely known.

A recent analysis shows that the top quark Yukawa coupling of any grand unified
theory leaves an imprint in the soft operators which leads to a violation
of individual lepton number violation, $L_i$ [12].
The argument can be summarized as follows.
All grand unified models must have a large superpotential interaction which
generates the top quark mass: $\lambda_t$.
Quark-lepton unification implies that this interaction breaks $\tau$ number.
Flavor mixing of the quarks implies  flavor mixing of the leptons, so this
interaction also breaks $\mu$ and $e$ number.
Hence the large top Yukawa coupling imprints the soft scalar operators with
$e, \mu$ and $\tau$ violation, leading to 1-loop weak scale contributions to
the
processes $\mu \rightarrow e \gamma$, $\tau \rightarrow \mu\gamma$,
$\mu \rightarrow 3e$ and $\mu\rightarrow e$ conversion.
Over much of the space of the soft parameters, the rates are within range of
future experiments which aim to push two orders of magnitude beneath present
experimental bounds [13].

In this letter we discuss how CP violating phenomena can probe supersymmetric
unification via the soft operators, and compute the size of the effects induced
by the top quark Yukawa coupling.
The most sensitive probes of CP violation beyond the standard model are the
electric dipole moments of the neutron $(d_n)$ and electron $(d_e)$.
This is because in the standard model the other CP observables, such as
$\epsilon, \epsilon', \sin 2\alpha$ and $\sin 2\beta$ are not
particularly suppressed, other than by small intergenerational mixing
angles.
On the other hand, a naive guess of $d_n \approx 10^{-25}$ e cm from the weak
scale is incorrect, because the chiral nature of the weak interaction
leads to vanishing electric dipole moments at both one and two loop order.
The actual prediction from the standard electroweak interaction is much
smaller,
of order $10^{-30}$ e cm [14].

{\bf 2.}

In models with weak scale supersymmetry, $d_n$ is generated
by the 1-loop diagrams of Figure 1.
We take the particle content of the low energy theory to be that of the minimal
supersymmetric standard model (MSSM), but allow for general soft
supersymmetry breaking scalar interactions:
$$
\eqalignno{
V_{soft} &= Q {\boldzeta}_U U^c H_2 + Q {\boldzeta}_D D^c H_1
+ L{\boldzeta}_E E^cH_1 + h.c.\cr
&+ Q^\dagger {\bf{m}}^2_Q Q + U^{c \dagger} {\bf{m}}^2_U U^c + D^{c \dagger}
{\bf{m}}^2_D D^c
+ L^\dagger {\bf{m}}^2_L L + E^{c \dagger} {\bf{m}}^2_EE^c&(2)\cr}
$$
where $Q$ and $L$ are squark and slepton doublets, and $U^c$, $D^c$ and $E^c$
are
squark and slepton singlets. The parameters ${\boldzeta}_U,$ etc,
are 3 $\times $ 3 matrices.
%We use lower case letters for the corresponding quarks and leptons.
Working in a basis in which the squark mass matrices $ {\bf{m}}^2_Q,
{\bf{m}}^2_U$ and
${\bf{m}}^2_D$ are diagonal,
the diagram of Figure 1 is proportional to the quantity $X$, where
$$
X = Im \left\{ V^T_{L_{1i}} (\zeta_D + \lambda_D \mu \tan \beta)_{ij}
V_{R_{j1}}\right\} I_{ij} (m^2_{Q_i}, m^2_{D_{j}})\eqno(3)
$$
where ${\boldlambda}_{U,D,E}$ are the quark and lepton Yukawa coupling
matrices.
In general there is an additional similar contribution involving internal up
squarks.
The parameter $\mu$ is the coefficient of the Higgs superpotential interaction,
$\mu H_1H_2$, while $\tan\beta=v_2/v_1$ is the ratio of vacuum expectation
values.
The matrix ${\bf V}_L$ (${\bf V}_R$) is the relative rotation between
left-handed
(right-handed) quarks and squarks to reach the mass eigenstate basis.
The function $I_{ij}$ results from the loop integral and depends on the squark
mass
eigenvalues.
Here and later we ignore the electroweak $D^2$ and supersymmetric
contributions to the squark and slepton masses.

Before studying unified theories we consider the MSSM.
The universal boundary condition allows us to choose ${\bf V}_R$ to be the
unit matrix at $M_P$.
Furthermore, a non-trivial ${\bf V}_R$ does not get generated
by renormalization group (RG) scaling.
Hence the quantity $X$, of equation (3), becomes
$$
X_{MSSM} = Im \left( V^T_{L_{1i}} (\zeta_D + \lambda_D \mu \tan \beta)_{i1}
\right)I_{i1}.\eqno(4)
$$
Keeping only the $RG$ scalings induced by the top Yukawa coupling, it is
possible to choose a basis in which
${\boldlambda}_U, {\boldzeta}_U$ and the squark masses are all real and
diagonal.
In this basis it is clear that there is no contribution to the electric dipole
moment from diagrams with internal up squarks.
The Yukawa couplings have the form
$$
W_{MSSM} = Q\overline{\boldlambda}_U U^c H_2 + Q {\boldlambda}_D D^c
H_1\eqno(5)
$$
where
$$
{\boldlambda}_D = {\bf{V}}^*\overline{\boldlambda}_D,\eqno(6)
$$
${\bf{V}}$ is the Kobayashi-Maskawa (KM) matrix, and $\overline{\boldlambda}_U$
and $\overline{\boldlambda}_D$ are real and diagonal.
For the MSSM, ${\bf V}_L={\bf V}$.
It may appear from (4) that a non-zero contribution to $X$ results when $i=3$,
with a size of order $Im(V^2_{td}) m \; \lambda_d I_{31}$, where $m$ is the
scale of supersymmetry breaking. However, this is incorrect.
A study of the $RG$ equations show that at low energies
$\zeta_{D_{31}} \propto \lambda_{D_{31}} \propto  V^*_{31}
\overline{\lambda}_{D_{11}}$.
Hence the quantity in (4) is proportion to $Im (V_{td} V^*_{td})=0$.

There is a deeper reason for this null result, which applies even in the case
of
large $\tan \beta$, when $RG$ scalings induced by $\lambda_b$
must also be kept.
The structure of CP violation in the MSSM with universal boundary conditions
is the same as for the standard model.
At any scale there is a single CP violating phase and it appears in ${\bf{V}}$.
Furthermore if any two eigenvalues of ${\boldlambda}_U$ or of ${\boldlambda}_D$
are equal then this phase can be removed.
The one loop diagram we have considered has a contribution which is
proportional to $m_d$.
Since it vanishes when $m_s=m_d$ and it is independent of $m_s$,
it is forced to vanish.

In the minimal supersymmetric $SU(5)$ grand unified model, the Yukawa
interactions are
$$
W_{SU(5)} = T\overline{\boldlambda}_U T \ H + T {\bf{P}}{\bf{V}}^*
\overline{\boldlambda}_D \overline{F} \; \overline{H}\eqno(7)
$$
where $T$ and $\overline{F}$ are 10 and $\overline{5}$ representations
of matter, $H$ and $\overline{H}$ are 5 and $\overline{5}$ Higgs
supermultiplets,
$\overline{\boldlambda}_U$ and $\overline{\boldlambda}_D$
are diagonal 3 $\times$ 3 Yukawa coupling matrices, ${\bf{V}}$ is the KM matrix
and
$\bf{P}$ is a diagonal phase matrix with two physical phases.
The reason for these two additional phases compared with the MSSM structure of
equations (5) and (6) is that if $\bf{P}$ is removed from the down couplings
by rephasing $T$, it appears in the up couplings.
In the MSSM it can be removed from the up couplings by relative notations of
$Q$ and
$U^c$, which is not possible in $SU(5)$ as these fields are unified into $T$.

Taking universal boundary conditions, ${\boldzeta}_D = A {\boldlambda}_D$,
${\boldzeta}_U = A {\boldlambda}_U$ and the squarks are degenerate at $M_P$.
Including $RG$ scaling effects induced by the top Yukawa coupling maintains the
diagonality of ${\boldlambda}_U, {\boldzeta}_U$
 and the squark masses.
The structure of (7) is also preserved, with the parameters becoming scale
dependent. In particular, no right-handed angles are generated:
${\bf{V}}_R$ of equation (3) is the unit matrix,
so that like the MSSM the quantity which determines the 1 loop
contribution to $d_n$ is that of equation (4).
Although the top quark Yukawa coupling causes some elements of ${\boldzeta}_D$
to scale in the $SU(5)$ theory,
it does not change their phase. This means that once $M_G$
is reached and the heavy states decouple, the phase matrix $\bf{P}$ can be
removed,
by rotating $Q$ and $U^c$ fields, in both the renormalizable and soft terms.
Thus the phases of $\bf{P}$ in the $SU(5)$ model do not lead to large
contributions to $d_n$ generated by the top Yukawa coupling.
The only physical phase of the low energy theory is in the KM matrix, and the
same argument as for the MSSM shows that there is no contribution to the
quantity $X$ at order $ Im (V^2_{td}) m \; \lambda_d I_{31}$.

{\bf 3.}

In the MSSM and $SU(5)$ models discussed above, we argued that the right
handed mixing angles, those appearing in ${\bf{V}}_R$ of equation 3,
are  zero.
This is because the universal boundary condition on the soft scalar
interactions allows these angles to be defined away at $M_P$,
and they are not generated by $RG$ scaling.
Another crucial feature that allows these angles to be defined away at $M_P$,
is that those theories both allow independent
rotations on the $Q_i$ fields and the $D^c_i$ fields.
In $SO(10)$ theories, however, $Q_i$ and $D^c_i$ are unified in the spinor
representation, $16_i$, so it is not possible to perform such relative
rotations.

In $SO(10)$ theories, the single Yukawa interaction $16 {\boldlambda} 16 \phi$,
where $\phi$ is a 10 dimensional Higgs multiplet, does not allow for any
intergeneration mixing, since the basis for the $16_i$ can
be chosen to make $\boldlambda$ diagonal.
We introduce a minimal $SO(10)$ model by the Yukawa interactions
$$
W_{SO(10)} = 16 {\boldlambda}_U 16 \; \phi_U + 16 {\boldlambda}_D 16 \; \phi_D.
\eqno(8)
$$
As always, we take the representation structure beneath $M_G$ to be that of
the MSSM, and we assume that the doublet $H_2$ lies
solely in $\phi_U$ and the doublet $H_1$ lies solely in $\phi_D$.
Thus ${\boldlambda}_U ({\boldlambda}_D$) is responsible for the up (down) quark
masses.

At $M_P$ we can choose a basis for the $16_i$ in which ${\boldlambda}_U$ is
real and diagonal. However, no further basis redefinitions are then possible.
Since ${\boldlambda}_D$ is symmetric it can be diagonalized by a single unitary
matrix
${\bf{U}}$: ${\boldlambda}_D = {\bf{U}}^* \overline{{\boldlambda}}_D
{\bf{U}}^\dagger
$. Hence we can write
$$
W_{SO(10)} = 16\overline{\boldlambda}_U 16 \; \phi_U + 16{\bf{U}}^*
\overline{\boldlambda}_D {\bf{U}}^\dagger 16 \; \phi_D\eqno(9)
$$
where $\overline{\boldlambda}_U$ and $\overline{\boldlambda}_D$ are diagonal.
The matrix ${\bf{U}}$ is a general $3 \times 3$ unitary matrix with 3 angles
and 6 phases. It can be written as
$$
{\bf{U}} = {\bf{P}}'^* {\bf{V}}{\bf{P}},\eqno(10)
$$
where ${\bf{V}}$ is the scale-dependent $KM$ matrix, and ${\bf{P}}$ and
${\bf{P}}'$ are diagonal phase matrices.

The $SO(10)$ RG equations can now be solved, keeping only the scalings induced
by the large top Yukawa coupling.
The matrices $\boldlambda_U, \boldzeta_U$
and the scalar mass matrix ${\bf{m}}^2$ remain diagonal.
It might be thought that at $M_G$,
when the superheavy states are decoupled, it will be possible to rotate the
$Q_i$ and $D^c_i$
states to remove the right-handed angles and extra phases, as was possible in
$SU(5)$.
In $SO(10)$ models this is not possible.
At $M_G$ the scalar mass squared matrix takes the diagonal form
$$
{\bf{m}}^2 = \pmatrix{ m^2_0&0&0\cr
                       0&m^2_0&0\cr
                       0&0& m^2_0 -I}\eqno(11)
$$
where $I$ is obtained from the RG equation for $m^2_3$ the third eigenvalue of
${\bf{m}}^2$
$$
I = {5\over 8\pi^2} \int^{M_P}_{M_G} (2 m^2_3 + m^2_\phi + A^2_3)
\lambda^2 dt\eqno(12)
$$
where $\lambda$ is the running top Yukawa coupling.
It is not possible to do a superfield rotation on $D^c$ to remove the
right-handed angles, because they reappear as a non-diagonal
${\bf{m}}^2_D$.
Theories which unify all quarks of a single generation will have right-handed
mixing
angles which are physical by virtue of large top Yukawa coupling effects.
At $M_G$, decoupling the superheavy states, we can write the Yukawa couplings
in the form
$$
W'_{SO(10)} = Q\overline{{\boldlambda}}_U U^c H_2 + Q ({\bf{V}}^*
\overline{\boldlambda}_D {\bf{P}}^{*2} {\bf{V}}^\dagger)D^cH_1\eqno(13)
$$
where ${\bf{V}}$ is the running $KM$ matrix. We could now redefine the $D^c_i$
super fields to put the matrix ${\bf{P}}^{*2} {\bf{V}}^\dagger$
in the soft operators.
This is logically preferable, since it is the presence of the soft operators
which allows the non-decoupling of these extra phases
and mixing angles.
However, for the calculation of $d_n$ it is preferable to remain in the present
basis,
which diagonalizes the squark mass matrix.
The presence of right-handed mixing angles means that there is no chiral
symmetry argument requiring $d_n \propto \lambda_d$,
the small down quark Yukawa coupling.
Similarly there is no argument that CP violation must vanish when two quark
mass
eigenvalues are  degenerate.
As we will see below, this allows very large contributions to
$X: X\approx  Im (V^2_{td}) m \; \lambda_b I_{33}$.

Our claim of such large effects is based on $I$ being a substantial
modification to $m^2_0$ for the third generation scalars.
Taking $A_3=0$,
and computing $I$ of equation (12) in the one loop insertion approximation
gives $I/m^2_0 \simeq 0.2 \lambda^2_G \ln {M_P\over M_G}$
where $\lambda_G = \lambda_t(M_G)$.
Present top quark mass data require $\lambda_G > 0.5$,
so $I/m^2_0 > 0.25.$
This should be considered a lower bound: in the integral of equation (12) the
effects of evolving
$\lambda$  and of allowing $A_3 \neq 0$
both lead to larger values for $I/m^2_0$.
Furthermore, the prediction for $m_b/m_{\tau}$ requires $\lambda_G$
larger than unity.
While this is relaxed if $\lambda_b$ is itself large, this requires $\tan\beta$
large (of order 50) which will enhance the dipole moment,
 as can be seen from equation (3).

{\bf 4.}

We now compute the contribution of the diagrams of Figure 1 to $d_n$,
and an analogous contribution to $d_e$, in the minimal $SO(10)$ theory.
In the superfield basis which diagonalizes the squark mass matrices, the Yukawa
interactions are given by (13) at $M_G$.
Modifications to this form will be induced by $\lambda_t$ scalings from
$M_G$ to $M_W$.
These effects are calculable, but they are not large and we ignore them for
simplicity. Hence
we take (12)  to apply also at the weak scale.

We have argued that at $M_G$ the third generation scalars are much lighter than
those of the first two generations.
This will also be true at the weak scale, although gluino contributions to
the squark masses could reduce the fractional difference.
In this letter will  therefore only calculate the contributions to $d_n$
arising from internal $b$ squarks.
The $\widetilde{b}_L$ and $\widetilde{b}_R$ squarks are degenerate at $M_G$.
Scalings induced by $\lambda_t$ will mean that at the weak scale
$\widetilde{b}_L$
is lighter than $\widetilde{b}_R$.
We ignore this effect for simplicity and take $\widetilde{b}_L$ and
$\widetilde{b}_{R}$
to be degenerate at the weak scale with mass $m_{\widetilde{b}}$.
We now rotate the quarks (not the squarks) to diagonalize the quark mass matrix
so that
we can compute the mixing matrices ${\bf{V}}_L$ and ${\bf{V}}_R$ of equation
(3).
We find ${\bf{V}}_L={\bf{V}}$ and
${\bf{V}}_R = {\bf{V}}{\bf{P}}^2$.
Hence we find
$$
X=Im (V^2_{td} P^2_{11} (\zeta_{D_{33}} + \lambda_{D_{33}} \mu \tan\beta))
I_{33}(m_{\widetilde{b}}^2, m_{\widetilde{b}}^2).\eqno(14)
$$

We write $\lambda_{D_{33}} = \lambda_b (V_{33} P_{33})^{*2}$
and $\zeta_{D_{33}} = A_b m_{\widetilde{b}} \lambda_b (V_{33} P_{33})^{*2}$,
where $A_b$ and $\lambda_b$ are real.
To relate $A_b$ to other A parameters of the theory would require solving the
$RG$ equation for $\boldzeta$.
This will be worth doing once the parameters of the MSSM are measured, but for
now we
 take $A_b$ to be an unknown parameter.
Hence
$$
X = \lambda_b A'_b  m_{\widetilde{b}} \; |V_{td}|^2 \; \sin \phi \;
I_{33} (m_{\widetilde{b}}^2, m_{\widetilde{b}}^2)\eqno(15)
$$
where $A'_b = A_b + { \mu\over m_{\widetilde{b}} } \tan\beta$, and we have set
$|V_{tb}| = 1$. The phase $\phi$ is a physical observable and is given by
$\phi = 2(\phi_{td} - \phi_{tb}) + 2(\phi_{11} - \phi_{33})$, where $\phi_{td}$
and $\phi_{tb}$ are the phases of $V_{td}$ and $V_{tb}$,
$P_{11} = e^{i \phi_{11}}$ and $P_{33} = e^{i \phi_{33}}$. Field rephasings can
change $\phi_{td}$, $\phi_{tb}$, $\phi_{11}$ and $\phi_{33}$,
but the phase $\phi$ is a physical observable.
Calculating the diagrams of Figure 1, with the gluino mass set equal to the
$b$ squark mass $\widetilde{m}_3 = m_{\widetilde{b}}$, we find
$$
d_n = e {2 \alpha_s \over 81 \pi} \; |V_{td}|^2 \; \sin \phi \;
{m_b\over \eta_b} \; A'_b \; {1\over m^2_{\widetilde{b}} }\eqno(16)
$$
where $\eta_b$ is the QCD correction required to run the $b$ quark mass up to
the scale of the superpartners.
To obtain this formula we have used the quark model result $d_n = 4 d_d/3$.

There is a similar contribution to the electric dipole moment of the electron,
$d_e$, coming from a diagram with internal bino and
tau sleptons.
In the minimal $SO(10)$ model, the slepton mass matrix at $M_G$ is given by
(12) and in this basis the lepton Yukawa matrix is given by
${\boldlambda}_E = {\bf{V}}^* \overline{{\boldlambda}}_D{\bf{P}}^{*2}
{\bf{V}}^\dagger$; it is identical to ${\boldlambda}_D$ given in (14).
Using an analogous set of assumptions,
and working in the approximation that the bino is a mass eigenstate,
one derives
$$
X_e = \lambda_{\tau} A'_{\tau}  \; |V_{td}|^2 \; \sin \phi \;
I_{33}(m^2_{\widetilde{\tau}}, m^2_{\widetilde{\tau}})
$$ where
$A'_{\tau} = A_{\tau} + {\mu\over m_{\widetilde{\tau}}} \tan\beta$.
Here $m_{\widetilde{\tau}}$ is the mass of both right and left-handed tau
scalars. Taking the bino degenerate with the tau slepton,
$\widetilde{m}_1=m_{\widetilde{\tau}}$, we find
$$
d_e = -e {\alpha\over 48 \pi c^2}  \; |V_{td}|^2 \; \sin \phi \;
m_{\tau} \; A'_{\tau} \; {1\over m^2_{\widetilde{\tau}}}\eqno(17)
$$
where $c^2= \cos^2\theta_W$.
Using the unification constraint on the gaugino masses:
$\widetilde{m}_3/\widetilde{m}_1 = \widetilde{\alpha}_3/\widetilde{\alpha}_1
\simeq 7$,
one can deduce that
$$
{d_n\over d_e}\simeq -0.4 {A'_b\over A'_{\tau}}\eqno(18)
$$
for the case that the scalars and gaugino in any loop are degenerate. The
relative minus sign is due to the fact that the product of the hypercharges of
left and right-handed sleptons is negative. We are unable to predict the
absolute sign of $d_n$ or $d_e$ since we do not know the sign of $A'_b$ or
$A'_{\tau}$.
Since we expect $A'_b$ and $A'_{\tau}$ to
be comparable, $d_n$ and $d_e$ are comparable in this limit.

The present experimental values for these quantities are
$$
d_n = (-30 \pm 50) \times 10^{-27} e \ cm\eqno(19)
$$
and
$$
d_e = (1.8 \pm 1.2 \pm 1.0) \times 10^{-27} e \ cm\eqno(20)
$$
from references [15] and [16] respectively.
Since the present experimental sensitivity to $d_e$ is
approximately 30 times greater than to $d_n$, equation (18) implies that, in
the degenerate mass case, $d_e$ provides a much stronger probe of the theory
than $d_n$.
If one takes
$\widetilde{m}_1 =m_{\widetilde{\tau}} = 100 $ GeV, and $\widetilde{m}_3 =
m_{\widetilde{b}}$ =
700 GeV, the central value predicted by (17), with $A'_{\tau} =1$, is already
excluded by
about a factor of 4, while the central prediction of (16), with $A'_b =1$, is
about an order of magnitude smaller than the experimental limit.
Consistency with data requires increasing the superpartner masses so that the
colored ones are above a TeV.
Thus the present data already suggest
that it is unlikely that the scalars and gauginos are degenerate.

Given the large hierarchy of $\widetilde{m}_3/\widetilde{m}_1 \simeq 7$,
it is reasonable to expect that the bino is lighter than the scalar tau.
In the limit that $\widetilde{m}_1 \ll m_{\widetilde{\tau}}$ we find
$$
d_e = -e {\alpha\over 8\pi c^2}  \; |V_{td}|^2 \; \sin \phi \;
m_{\tau} \; A'_{\tau} \;
{\widetilde{m}_1\over m^3_{\widetilde{\tau}}}\eqno(21)
$$
Evaluating (16) and (21) with representative masses we find:
$$
d_n = 40 \times 10^{-27} e \ cm
\left( {|V_{td}|^2 \over 10^{-4} } \right)
\left( {m_b/\eta_b\over 2.7 \; GeV} \right)
\left({ A'_b\over 1} \right)
\left({ 250 GeV\over m_{\widetilde{b}}} \right)^2
\left({ \sin \phi \over 0.5} \right) \eqno(22)
$$
and
$$
\eqalignno{
d_e = - 3.0 \times 10^{-27} e \ cm  &\left({|V_{td}|^2 \over 10^{-4}}\right)
                                      \left({m_{\tau}\over 1.78 \; GeV}\right)
                                      \left({A'_{\tau}\over 1}\right)\cr
                            &\left( {\widetilde{m}_1\over 35\; GeV}\right)
                           \left({200 \; GeV\over
m_{\widetilde{\tau}}}\right)^3
                               \left({ \sin \phi \over 0.5} \right)&(23)\cr}
$$
We have normalized these results to $\sin \phi = 0.5$. While the phase $\phi$
is not known, it is expected to be large. It has the same origin as the
physical phase in the Kobayashi-Maskawa matrix, which we know is not small.
With these parameter choices both predictions lie close to present
experimental limits.
Some regions of parameter space of the minimal SO(10) theory are already
excluded, for example those with light
scalar taus and large $\tan\beta \approx m_t/m_b$.
A complete numerical calculation must be done to determine precisely which
regions are already excluded, and how further improvements of the
experiments will constrain the theory.

To what extent will the above calculations for the minimal $SO(10)$
model apply to more general models?
The main assumption of the minimal $SO(10)$ model is that the quark and lepton
masses arise from the two renormalizable operators of equation 8.
However, in general there could be very many operators, including
non-renormalizable ones, contributing to the Yukawa interactions of the MSSM.
Thus, just beneath the GUT scale the Yukawa interactions need not have the
minimal $SO(10)$ from of equation (13).
While a quark basis can always be found to make ${\boldlambda}_U$ diagonal,
as shown in (13), the matrix ${\boldlambda}_D$ need not be symmetric,
so a general form for ${\boldlambda}_D$ should be taken:
${\boldlambda}_D = {\bf{V}} {\boldlambda}_D {\bf{V}}_R$.
Thus in equation (15) for $X$, and for $X_e$, one should replace
$|V_{td}|^2$ with $ |V_{td}V_{R_{td}}|$, and similarly in
equations (22) and (23) for $d_n$ and $d_e$. The definition of the phase $\phi$
will also change.

The precise values for $V_{R_{td}}$ are model dependent, indeed this becomes
the dominant uncertainty due to the unknown nature of the grand unified theory.
Nevertheless, the unified theory before symmetry breaking is left-right
symmetric, hence one expects $V_{R_{td}}$ to be comparable to $V_{td}$.
The predictions of this letter for the minimal $SO(10)$
model apply equally to arbitrary $SO(10)$ or $E_6$ models providing
$|V_{td}|^2$ is replaced by $|V_{td}V_{R_{td}}|$.
This does not alter the central value of the prediction, but does enlarge the
uncertainty of the prediction due to the model
dependence of the origin of quark and lepton masses.

{\bf 5.}

In this letter we have shown that supersymmetric theories, which unify the
quarks
and leptons of a generation into a single multiplet, lead to predictions for
$d_e$ and $d_n$ close to present experimental limits.
These predictions can be reliably computed in terms
of the parameters of the low energy supersymmetric theory.
The origin of this effect is the grand unified gauge symmetry and the top quark
Yukawa coupling. The unification of different particles into an irreducible
representation prevents basis rotations which remove flavor mixing angles.
The weak unification of ($u_L$ , $d_L$) into $Q$ prevents the removal of the
Kobayashi-Maskawa mixing matrix in the standard model. Similarly, the
unification $(Q, U^c, D^c)$ leads to a flavor mixing matrix in the right-handed
down quark sector, and the unifications  $(Q, U^c, E^c)$ and
$(Q, U^c, L)$ lead to flavor mixing matrices in the right-handed and
left-handed lepton sectors, respectively.

The super-rotations of $D^c, E^c$ and $L$ relative to $Q$ and $U^c$,
needed to diagonalize the fermion mass matrices, can only be performed
beneath $M_G$, where the grand unified symmetry is broken. However, at this
point the scalars are non-degenerate, and the flavor mixing reappears in the
scalar mass matrices. This non-degeneracy of the scalars, another
unavoidable feature of unification, occurs because the $D^c, E^c$ and $L$
particles also have interactions induced by the same large parameter,
$\lambda$, that induces the top quark mass.

Recently, large individual lepton number $(L_i)$
violating processes, such as $\mu \to e\gamma$, have also been proposed as a
signal for supersymmetric unification [12].
The origin of such signals is similar to that discussed here for the CP
violating  signals of $d_n$ and $d_e$, and a few brief comparisons can be made.
Both $L_i$ and $CP$
violating signals are based on the three assumptions of weak scale
supersymmetry, grand unification and supersymmetry breaking
operators present close to the Planck scale.
The $L_i$ violating signals are more general, in that they also apply to the
case
of $SU(5)$ and require only that top quark and tau lepton are unified.
For $SO(10)$ models the $L_i$ and $CP$ violating processes, very broadly
speaking, provide comparable signals.
For example, for slepton masses of 200 GeV, both the $\mu\to e$ conversion rate
and $d_e
$ are very close to present experimental sensitivities.
An important theoretical advantage of the CP violating signals is that as the
scale of supersymmetry
breaking, $m$, increases so the dipole moments scale as $ 1/m^2$,
while the $L_i$ violating processes have rates which
drop off more rapidly as $1/m^4$.
Thus, improvements of the experimental limits by the same factor would mean
that the
CP violating
signals would ultimately provide the most powerful probe of $SO(10)
$ theories.

We believe that a continued experimental search for electron and neutron
electric dipole moments will
provide a very powerful probe of supersymmetric $SO(10)$ unification.
Further numerical theoretical work is necessary to determine precisely how
present and future measurements will constrain the
parameter space of the low energy theory.
Suppose that supersymmetry is discovered, and the weak-scale parameters
associated with the superpartners
are measured. It will then be possible to compute the values of $d_n$ and
$d_e$, subject only to the mixing angle uncertainties.
If the moments are not seen at the predicted level, the theory will be
excluded.
This should be contrasted with the signals of proton decay and neutrino masses.
In general $SU(5)$ or $SO(10)$ models, the size of these signals cannot be
reliably computed,
and we can never foresee the exclusion of superunification on these grounds.

\vskip 4.5in

\newpage
\centerline{\bf References}

\begin{enumerate}
\item J. Pati and A. Salam, {\it Phys. Rev.} {\bf D8} 1240 (1973). \hskip 4in\\
H. Georgi and S. Glashow, {\it Phys. Rev. Lett. }  {\bf 32} 438. (1974).
\item H. Georgi in Particles and Fields, Proceedings of the APS Div. of
Particles
and Fields, ed. C. Carlson; H. Fritzsch and P.
Minkowski, {\it Ann. Phys. } {\bf 93} 193 (1975).
\item H. Georgi, H. Quinn and S. Weinberg, {\it Phys. Rev. Lett. } {\bf 33} 451
(1974).
\item S. Dimopoulos, S. Raby and F. Wilczek, {\it Phys. Rev. } {\bf D24}. 1681
(1981);
S. Dimopoulos and H. Georgi, {\it Nucl. Phys.} {\bf B193} 150 (1981);
L. Ibanez and G.G. Ross, {\it Phys. Lett. } {\bf 105B} 439
(1981).
\item L. Ibanez and G.G. Ross, {\it Phys. Lett.} {\bf 110B} 215 (1982); \\
L. Alvarez-Gaume, M. Claudson and M. Wise, {\it Nucl. Phys.} {\bf B207} 96
(1982).
\item M. Chanowitz, J. Ellis and M.K. Gaillard, {\it Nucl. Phys. } {\bf B128}
506 (1977).
\item G. Anderson, S. Dimopoulos, L.J. Hall, S. Raby and G. Starkman,
{\it Phys. Rev. } {\bf D49} 3660 (1994).
\item L.J. Hall, V. A. Kostelecky and S. Raby, {\it Nucl. Phys.} {\bf B267} 415
(1986).
\item A. Chamseddine, R. Arnowitt, and P. Nath, {\it Phys. Rev. Lett.}
 {\bf 49} 970 (1982);\\
R. Barbieri, S. Ferrara, and C. Savoy, {\it Phys. Lett.} {\bf B110} 343
(1982);\\
L. J. Hall, J. Lykken, and S. Weinberg, {\it Phys. Rev. } {\bf D27} 2359
(1983).
\item S. Dimopoulos and H. Georgi, Ref. 4; for a recent analysis see J. S.
Hagelin,
S. Kelley and T. Tanaka, {\it Nucl. Phys.} {\bf
B415} 293 (1994).
\item J. Ellis, S. Ferrara and D.V. Nanopoulos, {\it Phys. Lett.}
 {\bf 114B} 231 (1982);\\
W. Buchm\"uller and D. Wyler, {\it Phys. Lett.} {\bf 121B } 321 (1993);\\
J. Polchinski and M.B. Wise, {\it Phys. Lett.}{ \bf 125B} 393 (1983).
\item R. Barbieri and L.J. Hall, LBL-36022 (1994). To appear in {\it Phys.
Lett. B}.
\item R. Barbieri, L.J. Hall and A. Strumia, in preparation.
\item B. McKellar, S. Choudhury, $X-G.$ He and $S$. Pakvasa, {\it Phys. Lett. }
{\bf B197} 556 (1987), and reference therein.
\item I. S. Altarev et al., {\it Phys. Lett.} {\bf B276} 242 (1992).
\item E.D. Commins, S.B. Ross, D. Demille and C. Regan, {\it Phys. Rev. }
{\bf A}, Oct. 1994.
\end{enumerate}

\newpage

\centerline{\bf Figure Caption}

A Feynman diagram which provides a contribution to the neutron
electric dipole moment.

\end{document}